\documentclass[aps,pre,floats,twocolumn,showpacs,superscriptaddress]{revtex4}

\usepackage{graphicx,epsfig}
\usepackage{times}
\usepackage{graphics,dcolumn,bm,fleqn,epic,eepic,float}
\usepackage{amssymb,amsmath,multirow,rotate,color}
\begin{document}

\title{Synchronizability determined by coupling strengths and topology on Complex Networks}

\author{Jes{\'u}s G{\'o}mez-Garde\~{n}es}

\affiliation{Institute for Biocomputation and Physics of Complex
Systems (BIFI), University of Zaragoza, Zaragoza 50009, Spain}

\affiliation{Departamento de F\'{\i}sica de la Materia Condensada,
University of Zaragoza, Zaragoza E-50009, Spain}

\author{Yamir Moreno}

\affiliation{Institute for Biocomputation and Physics of Complex
Systems (BIFI), University of Zaragoza, Zaragoza 50009, Spain}

\author{Alex Arenas}

\affiliation{Departament d'Enginyeria Inform{\`a}tica i
  Matem{\`a}tiques, Universitat Rovira i Virgili, 43007 Tarragona,
  Spain}

\date{\today}

\begin{abstract}

We investigate in depth the synchronization of coupled oscillators on
top of complex networks with different degrees of heterogeneity within
the context of the Kuramoto model. In a previous paper
[Phys. Rev. Lett. 98, 034101 (2007)], we unveiled how for fixed
coupling strengths local patterns of synchronization emerge
differently in homogeneous and heterogeneous complex networks. Here,
we provide more evidence on this phenomenon extending the previous
work to networks that interpolate between homogeneous and
heterogeneous topologies. We also present new details on the path
towards synchronization for the evolution of clustering in the
synchronized patterns. Finally, we investigate the synchronization of
networks with modular structure and conclude that, in these cases,
local synchronization is first attained at the most internal level of
organization of modules, progressively evolving to the outer levels as
the coupling constant is increased. The present work introduces new
parameters that are proved to be useful for the characterization of
synchronization phenomena in complex networks.

\end{abstract}

\maketitle

\section{Introduction}

Studies on the emergence of collective and synchronized dynamics in
large ensembles of coupled units have been carried out since the
beginning of the nineties in different contexts and in a variety of
fields, ranging from biology, ecology, and semiconductor lasers, to
electronic circuits \cite{pik,k2,zanette}. Collective synchronized
dynamics has multiple applications in technology, and is a common
framework to investigate the crucial features in the emergence of
critical phenomena in natural systems. For instance, it is a relevant
issue to fully understand some diseases that appear as the result of a
sudden and undesirable synchronization of a large number of neuronal
units \cite{lglass}. Recently, synchronization phenomena have also
been proved to be helpful outside the traditional fields where it
applies, for instance, in sociology where it can be used to study the
mechanisms leading to the formation of social collective behaviors
\cite{vito1,vito2}.

Among the many models that have been proposed to address
synchronization phenomena, one of the most successful attempts to
understand them is due to Kuramoto
\cite{KuramotoLNP75,KuramotoBook84}, who capitalized on previous works
by Winfree \cite{WinfreeJTB67}, and proposed a model system of nearly
identical weakly coupled limit-cycle oscillators. The Kuramoto (KM)
mean field case corresponding to a uniform, all-to-all and sinusoidal
coupling is described by the equations of motion,
\begin{equation}
\dot{\theta}_{i}=\omega_{i}+\frac{K}{N}\sum_{j=1}^{N}
\sin{(\theta_{j}-\theta_{i})}
\hspace{0.5cm} (i=1,...,N)\;.
\label{eq:kuramodel}
\end{equation}
where the factor $1 / N$ is incorporated in order to ensure a good
behavior of the model in the thermodynamic limit,
$N\rightarrow\infty$, $\omega_i$ stands for the natural frequencies of
the oscillators, and $K$ is the coupling constant. Moreover, the
coherence of the population of $N$ oscillators is measured by the
complex order parameter,
\begin{equation}
r(t)\exp{({\mbox i}\phi(t))}=\frac{1}{N}\sum_{j=1}^{N}\exp{({\mbox
i}\theta_{j}(t))}\;,
\label{eq:kuraorderparam}
\end{equation}
where the modulus $0\le r(t) \le 1$ measures the phase coherence of
the population and $\phi(t)$ is the average phase. In what follows, we
will focus on the synchronization of coupled oscillators described by
the dynamics Eq.\ (\ref{eq:kuramodel}), because of its validity as an
approximation for a large number of nonlinear equations and its
ubiquity in the nonlinear literature \cite{conradrev}.

The KM approach to synchronization was a great breakthrough for the
understanding of the emergence of synchronization in large populations
of oscillators, in particular it presents a second-order phase
transition from incoherence to synchronization, in the order parameter
Eq.(\ref{eq:kuraorderparam}) for a critical value of the coupling
constant. However, a large amount of real systems do not show a
homogeneous pattern of interconnections among their parts
\cite{newmanrev,physrep} where the original KM assumptions apply.

Many real natural \cite{JeongNat01,SolePRSLB01}, social
\cite{NewmanPNAS01} and technological
\cite{FaloutsosCCR99,vespignanibook,WangPRE06} systems conform as
networks of nodes with connectivity patterns that diverge considerably
from homogeneity, and are usually characterized by a scale-free degree
distribution, $P(k)\sim k^{-\gamma}$ (the degree $k$ is the number of
connections of a node).  The study of processes taking place on top of
scale-free networks has led to reconsider classical results obtained
for regular lattices or random graphs due to the radical changes of
the system's dynamics when the heterogeneity of the connectivity
patterns can not be neglected
\cite{cnsw00,ceah00,ceah01,pv00,pv01,mpv02}. In this case one has to
deal with two sources of complexity, the nonlinear character of the
dynamics and the complex structures of the substrate, which are
usually entangled.  A contemporary effort to attack this entangled
problem was due to Watts and Strogatz, that in 1998, trying to
understand the synchronization of cricket chirps, which show a high
degree of coordination over long distances as though the insects where
``invisibly" connected, end up with a seminal paper \cite{WattsNat98}
about the small-world connectivity property. This work was the seed of
the modern theory of complex networks
\cite{newmanrev,physrep}. Nevertheless, the understanding of the
synchronization dynamics in complex networks still remains a
challenge.

In recent years, scientists have addressed the problem of
synchronization on complex networks capitalizing on the Master
Stability Function (MSF) formalism \cite{PecoraPRL98} which allows to
study the stability of the {\em fully synchronized state}
\cite{BarahonaPRL02,NishikawaPRL03,HongPRE04,ChavezPRL05,MotterPRE05,LeePRE05,DonettiPRL05,ZhouPRL06}.
The MSF is the result of a linear stability analysis for a completely
synchronized system. While the MSF approach is useful to get a first
insight into what is going on in the system as far as the stability of
the synchronized state is concerned, it tells nothing about how
synchronization is attained and whether or not the system under study
exhibits a transition similar to the original KM. To this end, one
must rely on numerical calculations and explore the {\em entire phase
diagram}. Surprisingly, there are only a few works that have dealt
with the study of the whole synchronization dynamics in specific
scenarios \cite{YamirEPL04,OhPRE05,McgrawPRE05,ArenasPRL06,
ArenasPhysD,usijbc} as compared with those where the MSF is used,
given that the onset of synchronization is reacher in its behavioral
repertoire than the state of complete synchronization.

In a previous work \cite{prljya}, we have shown how, for fixed
coupling strengths, local patterns of synchronization emerge
differently in homogeneous and heterogeneous complex networks, driving
the process towards a certain degree of global synchronization
following different paths. In this paper, we extend the previous work
to different topologies, even those with modular structure, and report
more results supporting the previous claim. First, we extend the
analysis carried out in \cite{prljya} to networks in which the degree
of heterogeneity can be tuned between the two limits of random
scale-free networks and random graphs with a Poisson degree
distribution. Second, in order to get further insights about the role
of the structural properties on the route towards complete
synchronization, we study the same dynamics on top of networks with a
non-random structure at the mesoscopic level, i.e., networks with
communities. The results support the usefulness of the tools developed
and highlight the relevance of synchronization phenomena to study in
detail the relationship between structure and function in complex
networks.

\section{KM model on complex networks}
\label{themodel}

Let us now focus on the paradigmatic Kuramoto model. In order to
manage with the KM on top of complex topologies we reformulate
eq. (\ref{eq:kuramodel}) to the form
\begin{equation}
\frac{d\theta_i}{dt}=\omega_i + \sum_{j}
\Lambda_{ij}A_{ij}\sin(\theta_j-\theta_i) \hspace{0.5cm} (i=1,...,N)\;,
\label{ks}
\end{equation}
where $\Lambda_{ij}$ is the coupling strength between pairs of
connected oscillators and $A_{ij}$ is the connectivity matrix ($A_{ij}=1$ if $i$ is linked
to $j$ and $0$ otherwise). The original Kuramoto model introduced
above assumed mean-field interactions so that $A_{ij}=1, \forall i\neq
j$ (all-to-all) and $\Lambda_{ij}=K/N, \forall i,j$.

The first problem when dealing with the KM in complex networks is the
definition of the dynamics. In the seminal paper by Kuramoto
\cite{KuramotoLNP75}, eq. (\ref{eq:kuramodel}), the coupling term in
the right hand side of eq. (\ref{ks}) is an intensive magnitude. The
dependence on the number of oscillators $N$ is avoided by choosing
$\Lambda_{ij}=K/N$. This prescription turns out to be
essential for the analysis of the system in the thermodynamic limit
$N\rightarrow \infty$. However, choosing $\Lambda_{ij}=K/N$
the dynamics of the KM in a complex network becomes dependent on
$N$. Therefore, in the thermodynamic limit, the coupling term tends to
zero except for those nodes with a degree that scales with $N$ \cite{note1}.

A second prescription consists of taking $\Lambda_{ij}=K/k_i$ (where
$k_i$ is the degree of node $i$) so that $\Lambda_{ij}$ is a weighted
interaction factor that also makes intensive the right hand side of Eq.
(\ref{ks}). This form has been used to solve the so-called {\em paradox of
heterogeneity} that states that the heterogeneity in the degree distribution,
which often reduces the average distance between nodes, may suppress
synchronization in networks of oscillators coupled symmetrically with uniform
coupling strength \cite{MotterPRE05}. One should consider this result
carefully because it refers to the stability of the {\em fully synchronized
state} (see below) not to the {\em whole evolution} of synchronization in the
network. More important, the inclusion of weights in the interaction strongly
affects the original KM dynamics in complex networks because it imposes a
dynamic homogeneity that could mask the real topological heterogeneity of the network.

Finally, the prescription $\Lambda_{ij}=K$
\cite{YamirEPL04,IchinomiyaPRE04,RestrepoPRE05}, which may seem more
appropriate, also presents some conceptual problems because the sum in the
right hand side of eq. (\ref{ks}) could eventually diverge in the
thermodynamic limit if synchronization is achieved. 
To our understanding, the most accurate interpretation of the KM dynamics in
complex networks should preserve the essential fact of treating the
heterogeneity of the network independently of the interaction dynamics, and at
the same time, should remain calculable in the thermodynamic limit. Taking
into account these factors, the interaction $\Lambda_{ij}$ in complex networks
should be inversely proportional to the largest degree of the system
$\Lambda_{ij}=K/k_{max}=\lambda$ keeping in this way the original
formulation of the KM valid in the thermodynamic limit (in SF networks
$k_{max}\sim N^{1/(\gamma-1)}$). In addition, the same order parameter,
eq. (\ref{eq:kuraorderparam}), can be used to describe the coherence of the
synchronized state.  Since $k_{max}$ is constant for a given network, the
physical meaning of this prescription is a re-scaling of the time units
involved in the dynamics. Note, however, that for a proper comparison of the
synchronizability of different complex networks, the global and local measures
of coherence should be represented according to their respective time
scales. Therefore, given two complex networks A and B with $k_{max}=k_A$ and
$k_{max}=k_B$ respectively, the comparison between observable must be done
for the same effective coupling
$K_A/k_A=K_B/k_B=\lambda$. With this formulation in mind
eq. (\ref{ks}) reduces to
\begin{equation}
\frac{d\theta_i}{dt}=\omega_i + \lambda \sum_{j}
A_{ij}\sin(\theta_j-\theta_i) \hspace{0.5cm} (i=1,...,N)\;,
\label{eq:kscn}
\end{equation}
independently of the specific topology of the network. This allow us
to study the dynamics of eq. (\ref{eq:kscn}) over different topologies
in order to compare the results and properly inspect the interplay
between topology and dynamics in what concerns to synchronization.

\section{Homogeneous {\em versus} heterogeneous topologies}
\label{sec:Kura-heterogeneous}

Recent results have shed light on the influence of the local
interactions' topology on the route to synchronization
\cite{McgrawPRE05,usijbc}. However, in these studies at least two
parameters (clustering and average path length) vary along the studied
family of networks. This paired evolution, although yielding an
interesting interplay between the two topological parameters, makes it
difficult to distinguish what effects were due to one or other
factors. Here, we would like to address first what is the influence of
heterogeneity, keeping the number of degrees of freedom to a minimum
for the comparison to be meaningful. The family of networks used in
the present section are comparable in their clustering, average
distance and correlations so that the only difference relies on the
degree distribution, ranging from a Poissonian type to a scale-free
distribution. Later on in this paper, we will relax these constraints
and study networks in which the main topological feature is given at
the mesoscopic scale, i.e., networks with community structure.

Therefore, let us first scrutinize and compare the synchronization
patterns in Erd\"os-R\'enyi (ER) and Scale-Free (SF) networks. For
this purpose we make use of the model proposed in \cite{presfer} that
allows a smooth interpolation between these two extremal
topologies. Besides, we introduce a new parameter to characterize the
synchronization paths to unravel their differences. The results reveal
that the synchronizability of these networks does depend on the
coupling between units, and hence, that general statements about their
synchronizability are eventually misleading. Moreover, we show that
even in the incoherent solution, $r=0$, the system is self-organizing
towards synchronization. We will analyze in detail how this
self-organization is attained.

The first numerical study about the onset of synchronization of
Kuramoto oscillators in SF networks \cite{YamirEPL04} revealed the
great propense of SF networks to synchronization, which is revealed by
a non-zero but very small critical value $\lambda_c$
\cite{note2}. Besides, it was observed that at the synchronized state,
$r=1$, hubs are extremely robust to perturbations since the recovery
time of a node as a function of its degree follows a power law with
exponent $-1$. However, how do SF networks compare with homogeneous
networks and what are the roots of the different behaviors observed?

We first concentrate on global synchronization for the Kuramoto model
Eq. (\ref{eq:kscn}). For this we follow the evolution of the order
parameter $r$, Eq. (\ref{eq:kuraorderparam}), as $\lambda$ increases,
to capture the global coherence of the synchronization in networks. We
will perform this analysis on the family of networks generated with
the model introduced in \cite{presfer}. This model generates a
one-parameter family of networks labeled by $\alpha\in[0,1]$.  The
parameter $\alpha$ measures the degree of heterogeneity of the final
networks so that $\alpha=0$ corresponds to the heterogeneous BA
network and $\alpha=1$ to homogeneous ER graphs. For intermediate
values of $\alpha$ one obtains networks that have been grown combining
both preferential attachment and homogeneous random linking so that
each mechanism is chosen with probabilities $(1-\alpha)$ and $\alpha$,
respectively.  It is worth stressing that the growth mechanism
preserves the total number of links, $N_{l}$, and nodes, $N$, for a
proper comparison between different values of $\alpha$. Specifically,
assuming the final size of the network to be $N$, the network is build
up starting from a fully connected core of $m_{0}$ nodes and a set
${\cal S}(0)$ of $N-m_{0}$ unconnected nodes. Then, at each time step,
a new node (not selected before) is chosen from $ {\cal S}(0)$ and
linked to $m$ other nodes. Each of the $m$ links is attached with
probability $\alpha$ to a randomly chosen node (avoiding self-
connections) from the whole set of $N-1$ remaining nodes and with
probability $(1-\alpha)$ following a linear preferential attachment
strategy \cite{doro}. After repeating this process $N-m_{0}$ times,
networks interpolating between the limiting cases of ER ($\alpha=1$)
and SF ($\alpha=0$) topologies are generated
\cite{presfer}. Furthermore, with this procedure, the degree of
heterogeneity of the grown networks varies smoothly between the two
limiting cases.

\begin{figure}[!t]
\begin{center}
\epsfig{file=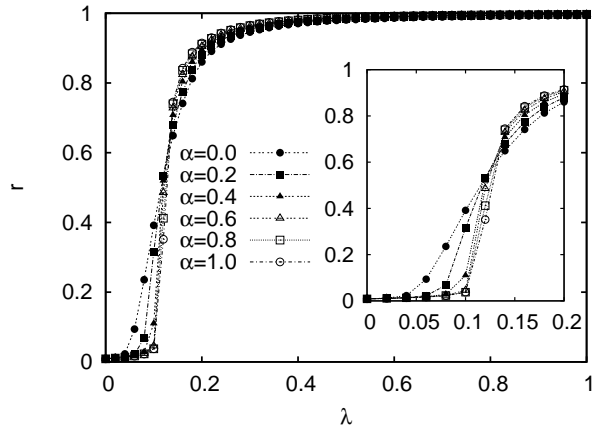,width=3.3in,angle=0,clip=1}
\end{center} 
\caption{Global synchronization curves $r(\lambda)$ for
  different network topologies labeled by $\alpha$ ($\alpha=0$
  corresponds to the BA limit and $\alpha=1$ to ER graphs). The inset
  shows the region where the onset of synchronization takes place. The
  network sizes are $N=10^4$ and $\langle k\rangle=6$ ($N_l=3\cdot
  10^4$) and were generated using the model introduced in
  \cite{presfer}}
\label{10000R}
\end{figure}

The curves $r(\lambda)$ for several network topologies ranging from ER
to SF are shown in Fig.\ref{10000R}. We have performed extensive
numerical simulations of eq. (\ref{eq:kscn}) for each network
substrate starting from $\lambda=0$ and increasing it up to
$\lambda=0.4$ with $\delta\lambda=0.02$. A large number (at least
$500$) of different network realizations and initial conditions were
considered for every value of $\lambda$ in order to obtain an accurate
phase diagram. The natural frequencies $\omega_i$ and the initial
values of $\theta_i$ were randomly drawn from a uniform distribution
in the interval $(-1/2,1/2)$ and $(-\pi,\pi)$, respectively.

Fig.\ref{10000R} reveals the differences in the critical behavior as a
function of the substrate heterogeneity. The global coherence of the
synchronized state, represented by $r$, shows that the onset of
synchronization first occurs for SF networks. As the network substrate
becomes more homogeneous the critical point $\lambda_c$ shifts to
larger values and the system seems to be less synchronizable. On the
other hand, it is also clear that the route to complete
synchronization, $r=1$, is faster for homogeneous networks. That is,
when $\lambda>\lambda_c(\alpha)$ the growth rate of $r$ increases with
$\alpha$. To inspect in depth the critical parameters of the system
dynamics we perform a finite size scaling analysis. This allows to
determine with precision the curve $\lambda_c(\alpha)$ and study the
critical behavior near the synchronization transition. We assume a
scaling relation of the form
\begin{equation}
r=N^{-\nu}f(N^{\beta}(\lambda-\lambda_c)),
\label{eqfss}
\end{equation}
where $f(x)$ is as usual a universal scaling function bounded as $x
\rightarrow \pm \infty$ and $\nu$ and $\beta$ are critical exponents
to be determined. The detailed analysis performed for both SF and ER
topologies shows that the critical value of the effective coupling,
$\lambda_c$, corresponds in scale-free networks to $\lambda_c^{SF} =
0.051$, and in random networks to $\lambda_c^{ER} = 0.122$,
accordingly with Fig.\ref{10000R}. In both cases, the transition
strongly recalls the classical transition of the original KM
\cite{KuramotoLNP75} with a critical exponent near $1/2$ for the SF
network \cite{YamirEPL04}. For intermediate values of $\alpha$, the
results show that the critical point shifts to larger values as the
degree of heterogeneity increases. They are shown in Table\
\ref{table1} together with some topological properties of the
networks. 

\begin{table}
\caption{\label{table1} Topological properties of the networks used in
  this work and critical points for the onset of synchronization
  obtained from a FSS analysis (Eq.\ (\ref{eqfss})). The topological
  quantities reported are the result of an average over 1000 network
  realizations. $\langle k \rangle=4 $ and $N=10^4$ have been set for
  all networks. Standard deviation of the mean values for $\lambda_c$
  is $\pm 2$ units in the last significant digit.}
\begin{ruledtabular}
\begin{tabular}{llll}
$\alpha$ & $ \langle k^2 \rangle $ & $k_{max}$ &
  $\lambda_c $\\
\hline
0.0 &	115.5 &	326.3 &	0.051\\
0.2 &	56.7 &	111.6 &	0.066\\	
0.4 &	44.9 &	47.7 &	0.088\\
0.6 &	41.1 &	25.6 &	0.103\\
0.8 &	39.6 &	16.8 &	0.108\\
1.0 &	39.0 &	14.8 &	0.122
\end{tabular}
\end{ruledtabular}
\end{table}

The differences between ER and SF topologies observed when looking at
global patterns of synchronization motivate a more detailed study of
the synchronization onset for both topologies. The original work by
Kuramoto pointed out that at the onset of synchronization small
clusters of locked oscillators emerge and that the recruitment of more
oscillators into these clusters as the coupling is increased makes it
larger the global coherence $r$ of the system. Obviously the emergence
of these clusters would depend on the underlying topology which drives
the possible configurations that locked oscillators would eventually
form. To see how this initial coherence is achieved we
propose a new order parameter, $r_{link}$. This parameter measures the
local construction of the synchronization patterns \cite{note3} and
allows for the exploration of how global synchronization is
attained. We define
\begin{equation}
r_{link}=\frac{1}{2N_l}\sum_{i}\sum_{j\in \Gamma_i}\left |\lim_{\Delta
  t\rightarrow\infty}\frac{1}{\Delta t}\int_{t_r}^{t_r+\Delta
  t}e^{i\left[\theta_i(t) -\theta_j(t)\right]}dt\right |,
 \label{r_link}
\end{equation}
being $\Gamma_i$ the set of neighbors of node $i$. The parameter
$r_{link}$ measures the fraction of all possible links that are
synchronized in the network. The averaging time $\Delta t$ should be
taken large enough in order to obtain good measures of the degree of
coherence between each pair of physically connected nodes. Besides,
$r_{link}$ is computed after the system relaxes at some large time
$t_r$. Note that in the limit of all-to-all coupling the information
provided by $r_{link}$ is exactly the same that the one provided by
$r$ because in this case $r_{link}\propto r^2$. Therefore, no
additional information would be provided by this new parameter in the
all-to-all case. Here, however, it turns out to be the key parameter
to characterize how synchronization emerges at a local scale.

\begin{figure}[!t]
\epsfig{file=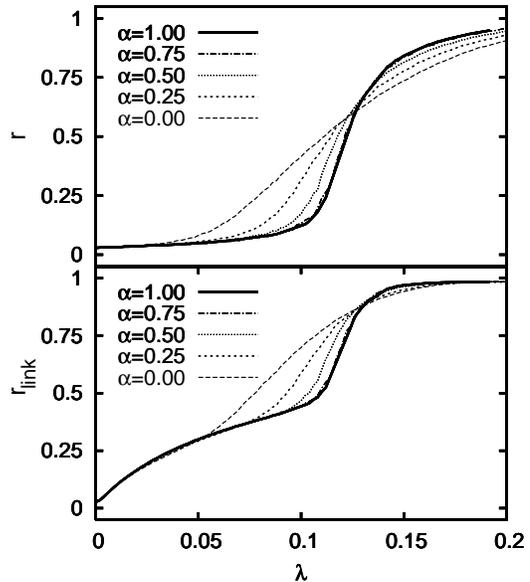,width=2.8in,angle=0,clip=1}
\caption{Evolution of the control parameters $r$ and $r_{link}$ as a
  function of the coupling strength for networks generated with the
  model introduced in \cite{presfer}, corresponding to $\alpha=0.0$,
  $0.25$, $0.5$, $0.75$ and $1.0$. The size of the networks is
  $N=10^3$ and their average degree is $\langle k \rangle=6$. The
  exponent of the SF networks increases from $\gamma=3$
  ($\alpha=0$).}
\label{R}
\end{figure}

In Fig.\ref{R} we represent the evolution of both order parameters,
$r$ and $r_{link}$, as a function of the coupling strength $\lambda$
for several values of $\alpha$. The behavior of $r_{link}$ shows a
change in synchronizability between ER and SF and provides additional
information to that reported by $r$. Interestingly, the nonzero values
of $r_{link}$ for $\lambda\leq\lambda_c$ indicate the existence of
some local synchronization patterns even in the regime of global
incoherence ($r \approx 0$). Right at the onset of synchronization for
the SF network limit, its $r_{link}$ value deviates from that of the
ER recovering the known result about the synchronization of SF
networks for lower values of the coupling. In this region, while the
synchronization patterns continue to grow for the ER network at the
same rate, the formation of locally synchronized structures occurs at
a faster rate in the SF network. Finally, when the incoherent solution
in the ER network destabilizes, the growing in its synchronization
pattern increases drastically up to values of $r_{link}$ comparable to
those obtained in SF networks and even higher. For intermediate values
of $\alpha$, the results show that the effect of varying the
heterogeneity of the underlying network is twofold. On one hand, the
more heterogeneous the network is, the smaller the values of $\lambda$
needed for the onset of synchronization. Conversely, the increase in
the degree of heterogeneity results in larger values of $\lambda$ in
order to achieve complete synchronization. In short, as the
heterogeneity is increased, the onset of synchronization is
anticipated, but at the same time, the appearance of the fully
synchronized state is delayed.

These results undoubtedly point out that statements about
synchronizability are dependent on the coupling strength value. To
shed new light on this phenomenon, we have studied the characteristics
of the synchronization patterns along the evolution of
$r_{link}$. Following the usual picture, synchronization patterns are
formed by pairs of oscillators, physically connected, whose phase
difference in the stationary state tends to zero. In order to
determine which pairs of nodes are truly synchronized we should
determine the coherence of their dynamics. Note that eq.(\ref{r_link})
is the average dynamical coherence between every pair of linked nodes
and then the synchronization degree of every pair of connected
oscillators can be written in terms of a symmetric matrix
\begin{equation}
{\cal D}_{ij} =A_{ij}\left |\lim_{\Delta t\rightarrow\infty}\frac{1}{\Delta
  t}\int_{t_r}^{t_r+\Delta t}e^{i\left[\theta_i(t)-\theta_j(t)\right]}dt\right
  |\;.
\label{eq:Dij}
\end{equation}
Then one has to analyze each matrix term $D_{ij}$ in order to label a
link $(i,j)$ as synchronized or not. As introduced above, from the
computation of $r_{link}$ one determines the fraction of physical
links that are synchronized so that one would expect that
$2r_{link}\cdot N_l$ elements of the matrix ${\bf {\cal D}}$ are
$D_{ij}= 1$, while the remaining elements are $D_{ij}= 0$. However,
this is not the real situation since the network dynamics is not well
defined in terms of a fully synchronized cluster and a set of
completely incoherent oscillators.  On the other hand, the worst
scenario would be found if there were $2N_l$ elements of matrix ${\bf
{\cal D}}$ so that $D_{ij}=r_{link}$, implying that all the physically
connected pairs are equally synchronized and hence the parameter
$r_{link}$ could not be interpreted as the fraction of links that are
dynamically coherent and no information about the topological patterns
of synchronization could be extracted from matrix ${\bf {\cal D}}$.
The situation found is not as simple as the former possibility and not
so dramatic as the latter. The contributions $D_{ij}$ of the $N_{l}$
elements of matrix ${\bf {\cal D}}$ that correspond to physical links
can be ordered from the highest to the lowest one. We have checked
that for two situations, corresponding to the onset of synchronization
($\lambda=0.05$) and when high global coherence ($\lambda=0.13$) is
observed for a SF network, synchronized links can be clearly
identified.  For the onset of synchronization, a subset of nearly $20
\%$ of links displaying coherent dynamics with high degree of
synchronization, ${\cal D}_{ij}>0.8$, is well separated from the
behavior of the remaining links as a dramatic decrease of $D_{ij}$
takes place.  In this sense, it is clear that the dynamics of a $20\%$
of the possible pairs can be regarded as synchronized which is in
agreement with the obtained value $r_{link}=0.25$ for $\lambda=0.05$
and support that although macroscopic coherence is not observed
($r\simeq0$ at this point) the system is seen to walk towards it. For
$\lambda=0.13$ ($r_{link}\simeq 0.82$) a plateau of nearly $75 \%$ of
links is observed, thus revealing the high degree of global coherence,
$r\simeq 0.7$, at this point. Therefore, the shape of the ranked
${\cal D}_{ij}$ curves confirm that $r_{link}$ gives the fraction of
synchronized links and thus the latter allows to obtain information
about synchronized patterns from ${\bf {\cal D}}$.

To determine exactly which pairs of nodes are regarded as
synchronized, the matrix ${\cal D}$ is filtered using a threshold $T$
such that the fraction of synchronized pairs equals $r_{link}$. In
this way $T$ is a moving threshold so that if ${\cal D}_{ij}>T$
oscillators $i$ and $j$ are considered synchronized. The value of $T$
depends on the particular realization and is determined by means of an
iterative scheme starting from $T=1$. Decreasing it with $\delta
T=0.01$ one computes the amount of links that fulfills the
condition. In this way, the value of $T$ progressively decreases and
more pairs of oscillators are chosen. The process lasts until $T$ is
such that the fraction of chosen links is equal to the desired value
$r_{link}$ previously computed from ${\cal D}$. Finally, when the
synchronized links are identified the clusters of synchronized nodes
are reconstructed.

\begin{figure}[!t]
\epsfig{file=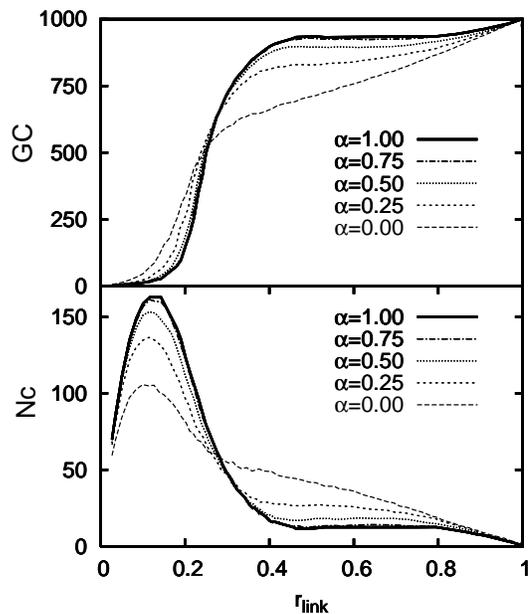,width=2.8in,angle=0,clip=1}
\caption{Evolution of the number of synchronized clusters
  $N_c$ and the synchronized giant component size $GC$ as a function of
  $r_{link}$ for the the different topologies
  considered. Small values of $r_{link}$ correspond to values of $\lambda$ for
  which $r\approx 0$. Despite $r$ being vanishing and hence no global
  synchronization is yet attained, a significant number of clusters show
  up. This indicates that for any $\lambda>0$ the system self-organizes
  towards macroscopic synchronization. The network parameters are as in Fig.\
  \ref{R}.}
\label{pattern}
\end{figure}

\begin{figure*}[!t]
\begin{center}
\epsfig{file=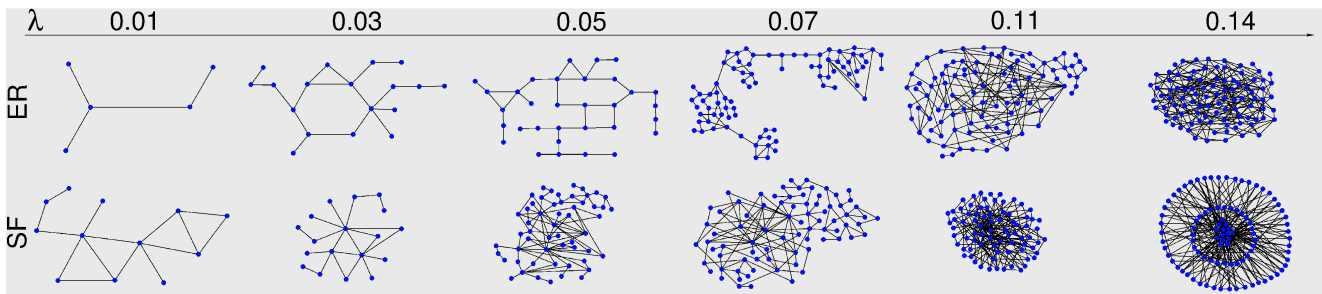,height=1.5in,angle=0,clip=1}
\end{center}
  \caption{Giant synchronized components for several values of
  $\lambda$ in the two limiting cases of the different topologies
  studied (ER and SF). The size of the underlying networks is small
  ($N=100$ nodes), in order to have a sizeable picture of the
  system. Note that for the SF case links and nodes are incorporated
  together to the $GC$, while for the ER network, what is added are
  links between nodes already belonging to the $GC$.}
\label{nets}
\end{figure*}

Figure \ref{pattern} represents the number of synchronized clusters
and the size of the giant component ($GC$) as a function of $r_{link}$
for the same values of $\alpha$ used in Fig.\ \ref{R}. The local
information extracted from it points to a novel feature of the
synchronization process that is not possible to derive from
Figs.\ref{10000R} and \ref{R}, and that is unexpected. The emergence
of clusters of synchronized pairs of oscillators (links) in the
networks shows that for values of $\lambda$ for which the incoherent
solution $r=0$ is stable, the networks have developed a largest
cluster of synchronized pairs of oscillators involving $50\%$ of the
nodes, and an equal number of smaller synchronization clusters. From
this point on, the behavior of both $GC$ and $N_c$ depends on the
specific value of $\alpha$. When heterogeneity dominates, the GC grows
and the number of smaller clusters goes down, whereas for less
heterogeneous networks the growth of $GC$ is more abrupt and nodes are
incorporated to it more faster. Moreover, the results highlight the
fact that although heterogeneous networks exhibit more coherence in
terms of $r$ and $r_{link}$, the microscopic evolution of the
synchronization patterns is faster in homogeneous networks, being
these networks far more locally synchronizable than the heterogeneous
ones once $\lambda>\lambda_c$.

The observed differences in the behavior at a local scale are rooted
in the growth of the $GC$. For homogeneous topologies, many small
clusters of synchronized pairs of oscillators (note in
Fig.\ref{pattern} the large number of clusters formed when a $15 \%$
of the links are synchronized) merge together to form a GC when the
effective coupling is increased. This coalescence of many small
clusters results in a giant component made up of almost the size of
the system once the incoherent state destabilizes. On the other hand,
for heterogeneous graphs, the growth of the giant component is more
smooth and the oscillators form new pairs starting from a core made up
of half the nodes of the network. That is, in one case (ER-like
networks), almost all the nodes of the network takes part of the giant
component from the beginning and latter on, when $\lambda$ is
increased, what is added to the $GC$ are the links among these nodes
that were missing in the original cluster of synchronized nodes. For
SF-like networks, the mechanism is the opposite. Nodes are added to
the $GC$ {\em together} with most of their links, resulting in a
growth of $r_{link}$ much slower than for the homogeneous topologies.

\begin{figure}[!b]
\begin{center}
\epsfig{file=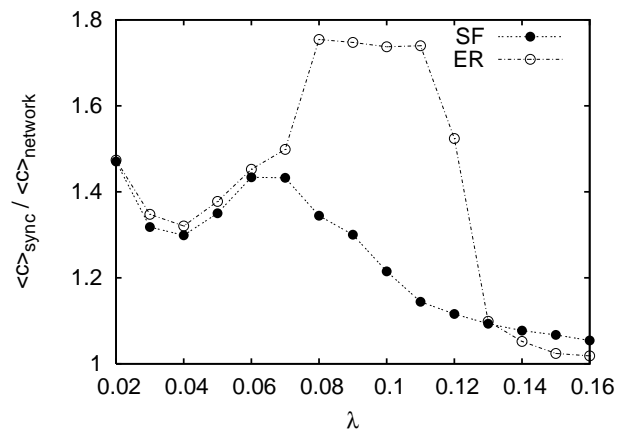,width=2.3in,angle=-90,clip=1}
\end{center} 
 \caption{Evolution of the ratio between the clustering coefficient of
 the giant synchronized cluster, $\langle c_{sync}\rangle$, and that
 of the substrate network $\langle c_{network}\rangle$, as a function
 of $\lambda$ for the two limiting cases of BA and ER
 networks. Network parameters are those used in figure \ref{R}.}
\label{clust}
\end{figure}

The above picture is confirmed in Fig.\ref{nets}, where we have
represented the evolution of the local synchronization patterns of the
giant components in ER and SF networks for several values of $\lambda$
\cite{note4}. It is clear that when $r\simeq 0$ the two networks
follow different paths toward synchronization. In particular, the
giant component for the SF network seems to retain the topological
features of the substrate network, while this is not the case for the
ER network (for instance, the small-world property is clearly
lacking).

This study about the patterns of self-organization towards
synchronization reveals that the quantitative difference about the
macroscopic behavior, shown by the computation of the evolution of the
global coherence $r$ for ER and SF networks, has its roots on a
qualitatively different route at the microscopic level of
description. The use of the new parameter $r_{link}$ which involves
the computation of the degree of coherence between each pair of linked
nodes is a useful tool for describing such differences. Moreover, the
results suggest that the degree of heterogeneity of the network is the
key ingredient to explain the two different routes observed. The
technique developed to extract the synchronization patterns allows the
analysis of the topological features of such clusters of nodes. We can
compute the average measures of relevant quantities such as the
clustering coefficient or the degree distribution, and see how these
magnitudes evolve from the uncoupled limit, where no synchronization
occurs, to the coherent regime where the synchronized network
coincides with the underlying substrate. It is then relevant to
explore the regions where the onset of synchronization takes place and
characterize topologically these emergent synchronized clusters.

In Fig.\ref{clust} the evolution of the average clustering coefficient
$\langle c_{sync}\rangle$ of the giant synchronized cluster referred
to $\langle c_{network}\rangle$ in the underlying network, is plotted
as a function of $\lambda$ for both the BA and ER networks. It is
worth mentioning that the results depicted in the figure have been
computed taking into account that nodes with degree 1 does not
contribute to the clustering coefficient of the $GC$, as $c$ is not
properly defined for these nodes. The results are illustrative of the
local organization of synchronized nodes. The figure shows that for
both topologies the clustering decreases as the coupling is increased
beyond their respective $\lambda_c$ or, in other words, as the giant
component grows by the addition of new synchronized pairs of
nodes. However, the effects of the two different routes to complete
synchronization observed for ER and SF networks are well appreciated
from the results. For the heterogeneous network there is a smooth
decrease of the clustering coefficient for $\lambda>\lambda_c^{SF}$
and the effects of the emergence of global coherence are not dramatic
in what refers to the behavior of $\langle c_{sync}\rangle$. This is
because in this case the giant component mainly grows by recruiting
new synchronized nodes and their links. On the other hand, for the ER
graph the behavior observed for $\lambda<\lambda_c^{ER}$, i.e. when no
macroscopic coherence is observed, is interrupted by a sudden jump
near its critical value. In fact, for $\lambda>\lambda_c^{ER}$ the
clustering of the synchronized cluster quickly approaches the value of
$\langle c\rangle$ of the substrate network. This effect becomes clear
if one has in mind the coalescence of small clusters, which happens
around the critical point for ER graphs. In fact, taking into account
the giant synchronized component on ER for $\lambda<\lambda_c^{ER}$,
implies to consider one of the several disjoint synchronized clusters
of similar sizes that are in this region. Moreover, the coalescence
process leads to the formation of a giant cluster that contains almost
all the nodes of the network (see Fig.\ \ref{pattern}), but a number
of links significantly smaller. Hence, when the clusters collapse into
a much larger one, the topological features change dramatically as
observed from the evolution of the clustering coefficient.

\begin{figure}
\begin{center}
\epsfig{file=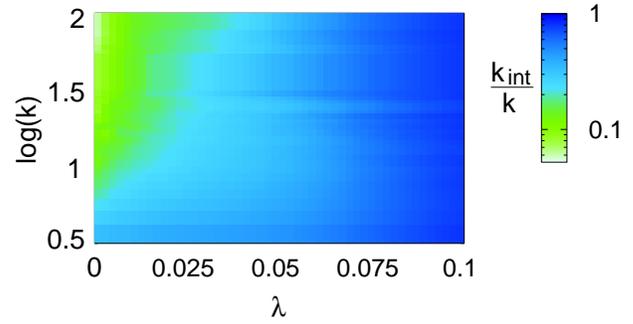,width=3.4in,angle=0,clip=1}
\end{center} 
 \caption{(color online) The plot shows the fraction of links that a
  node with degree $k$ belonging to the synchronized cluster shares
  with other nodes of the same synchronized cluster. This fraction
  $k_{int}/k$ is plotted as a function of $k$ and $\lambda$. The
  figure shows how the hubs progressively incorporate their neighbors
  to the synchronized component as $\lambda$ grows. The network is SF
  with parameters as those used in Fig.\ref{R} and $\alpha=0$.}
\label{ki}
\end{figure}

All the results reported above point out that the ultimate reason
behind the two different routes to complete synchronization is the
heterogeneous character of the SF network and the role played by the
hubs.  The natural cohesion that hubs provide to SF network prevents
the existence of independent macroscopic clusters of synchrony as
occurs for ER networks. It is then interesting to study how these hubs
participate in the formation of the final synchronized state. For
this, we first study the evolution with $\lambda$ of the composition
of the synchronized cluster in terms of the degree of its
components. In \cite{prljya}, we reported the probability that a node
with degree $k$ belongs to the giant synchronized cluster as a
function of its degree $k$ and the coupling $\lambda$ for the SF
network. This probability turns out to be an increasing function of
$k$ for every value of $\lambda$ and therefore the more connected a
node is, the more likely it takes part in the cluster of synchronized
links. In particular, the results confirm the hypothesis made above
that the hubs participate from the very beginning on the formation of
the synchronized cluster. A similar result was obtained in
\cite{ZhouChaos06}, where Zhou and Kurths studied the hierarchical
organization in complex networks, using the MSF and a mean-field
approach in the weak coupling limit.

The above characterization of the synchronized cluster in terms of the
degree of its component can be completed studying their effective
degree, $k_{int}$. The effective degree of a synchronized node is the
number of links it shares with other nodes belonging to the same
synchronized cluster. Obviously, at the complete synchronized regime a
node with degree $k$ will have $k_{int}=k$. We have plotted in
Fig.\ref{ki} the quantity $k_{int}/k$ (the fraction of links that a
node has with synchronized neighbors) as a function of $\lambda$ and
the degree $k$ of the nodes ($\alpha=0$). The results reveal that
although hubs are the first to take part of the synchronized cluster,
their neighbors are progressively incorporated to the cluster as
$\lambda$ grows. Besides, if a node with small $k$ is synchronized the
probability that its neighbors are also synchronized grows very fast
with $\lambda$ which is an effect of the network topology.  These
results further support the statement about the essential role played
by the hubs in the recruitment of oscillators into the synchronized
group and in the emergence of complete synchronization in SF networks.

\section{Synchronization in structured networks}
\label{sec:Kura-community}

In light of the results of the above section we have extended the
study beyond unstructured networks to structured or modular
networks. This is a limiting situation in which the local structure
may greatly affect the dynamics, irrespective of whether or not we deal
with homogeneous or heterogeneous networks and then they constitute a
perfect framework for testing the new order parameter $r_{link}$
introduced in the last section.

Many complex networks in nature are modular, i.e. composed of certain
subgraphs with differentiated internal and external connectivity that
form communities \cite{physrep,alexcomm}. The use of modular networks
where a proper comparison in synchronizability can be performed (same
number of nodes and links) restricts us to the consideration of
synthetic structured networks. To this end, we make use of a common
benchmark of random network with community structure, first proposed
by Newman \cite{NewmanPRE04} considering one hierarchical level and
later extended to several hierarchical levels \cite{ArenasPRL06,
ArenasPhysD}.
   
The modular network structure we build is as follows: in a set of $N$ nodes,
we prescribe $n$ compartments that will represent our first community
organizational level, and $m$ compartments, each one embedding four different
compartments of the first level, that define the second organizational level
of the network. The internal degree of nodes at the first level $z_{in_1}$ and
the internal degree of nodes at the second level $z_{in_2}$ keep an average
degree $z_{in_1}+z_{in_2}+z_{out}=\langle k\rangle$ so that these networks are
strictly homogeneous in the sense of the degree distribution ,
$P(k)=\delta(k-\langle k\rangle)$. Networks with two hierarchical levels are
labeled as $z_{in_1}$ - $z_{in_2}$, e.g. a network with $i$-$j$ means $i$
links with the nodes of its first hierarchical community level (more
internal), $j$ links with the rest of communities that form the second
hierarchical level (more external) and $(\langle k\rangle-i-j)$ links with any
community of the rest of the network.

\begin{figure}[!t]
\begin{center}
\epsfig{file=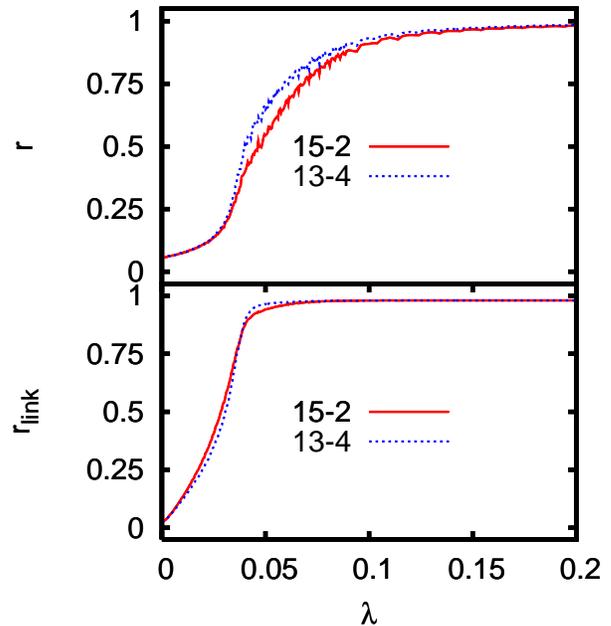,width=3.4in,angle=-90,clip=1}
\end{center} 
 \caption{(color online) Evolution of $r$ (top) and $r_{link}$
  (bottom) as a function of $\lambda$ for structured modular
  networks. The networks are synthetically built with an {\em a
  priori} community structure. The network size is 256 nodes and the
  number of links is 4608. We prescribe 16 compartments that will
  represent our first community organizational level, and four
  compartments each one embedding four different compartments of the
  above first level, that define the second organizational level of
  the network.  Each node has 18 links distributed between its first
  community level, the second, and the whole network at random. The
  network 13-4 has 13 internal connections in its first hierarchical
  level, 4 external connections in its second hierarchical level, and
  1 connection with any other community in the network. The generation
  of the 15-2 structure is equivalent. The curves show that although
  13-4 has always a better global synchronization, 15-2 has better
  local synchronization as shown by $r_{link}$.}
\label{Communities}
\end{figure}

Synchronization processes on top of modular networks of this type have
been recently studied as a mechanism for community detection
\cite{ArenasPRL06,vitosync}. In \cite{ArenasPRL06}, the authors
studied the situation in which starting from a set of homogeneous (in
terms of the natural frequencies) Kuramoto oscillators with different
initial conditions the system evolves after a transient of time to the
synchronized state. It was shown that the community structure is
progressively unveiled at the same time the system's dynamics evolves
toward the coherent state and the synchronization is attained. In
particular, the nodes belonging to the first community level are the
first to get synchronized, subsequently the second level nodes achieve
the frequency entrainment and finally the whole system shows global
synchronization.

\begin{figure}[!t]
\begin{center}
\epsfig{file=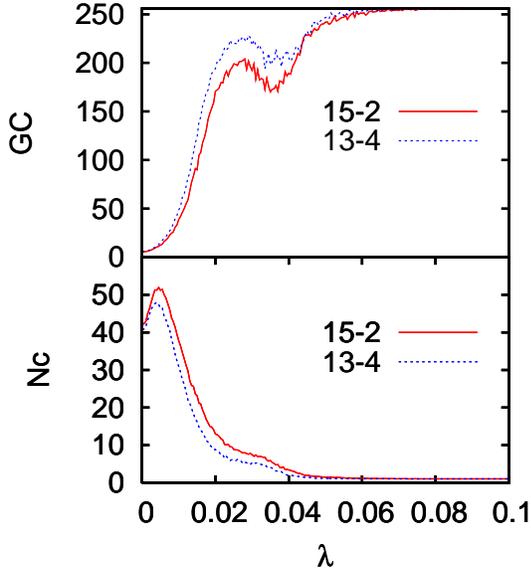,width=3.1in,angle=-90,clip=1}
\end{center}
\caption{(color online) Size of the largest synchronized cluster $GC$
  (top) and number of clusters $N_c$ (bottom) for the same networks of
  Fig.\ \ref{Communities}. See the text for details.}
\label{gc_comm}
\end{figure}

\begin{figure}[!t]
\begin{center}
\epsfig{file=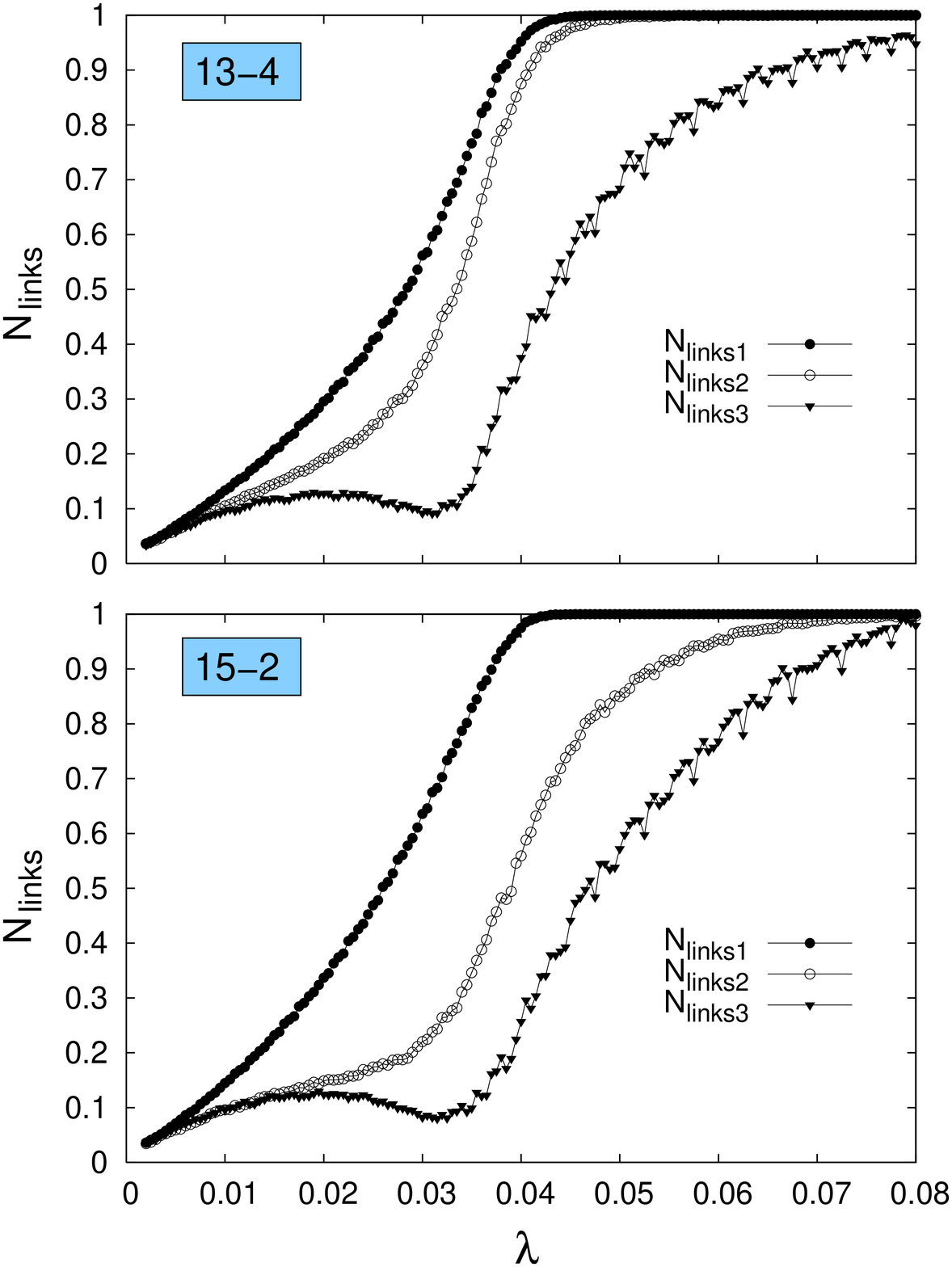,width=2.7in,angle=0,clip=1}
\end{center}
\caption{Number of links ($N_{links}$) that connect synchronized nodes
  in each of the three levels in the community hierarchy of the
  networks ($1$ means inner layer). The numbers are normalized by the total
  number of links at each level in each network.}
\label{nlinks}
\end{figure}

Here we adopt a different perspective since we will consider as previously a
set of non-identical Kuramoto oscillators with random assignment of natural
frequencies and hence the final degree of system's synchronization will depend
on the strength of the coupling. It is then interesting to study how the
degree of synchronization evolves as a function of $\lambda$ and whether the
coherence between nodes is progressively distributed following the hierarchy
imposed by the underlying topology. For this, we make use of the order
parameters $r$, eq. (\ref{eq:kuraorderparam}), and $r_{link}$, eq
(\ref{r_link}), to characterize the synchronization transition on two slightly
different modular networks with two well defined hierarchical levels, $13-4$
and $15-2$, being this difference the cohesion of the internal community core,
13 links out of 15 possible neighbors or 15 links (i.e., all-to-all) at the
most internal level. Both networks have $N=256$ and $\langle
k\rangle=18$. Fig.\ref{Communities} shows the results for both kinds of
networks revealing that the path towards synchronization as a function of the
interaction is again affected by the structure. They also show that the
information provided by $r_{link}$ is essential to unveil the synchronization
process. While the global synchronization parameter $r$ is reflecting that the
$13-4$ structure globally synchronizes always better, $r_{link}$ tells us
again about the local synchronization. It shows that local synchronization is
indeed favored in the $15-2$ structure since $r_{link}$ is larger for this
topology for small values of $\lambda$ where the system is locally forming
synchronized clusters. This result, not captured by the macroscopic indicator
$r$, is expected since the internal cohesion of communities at the first
hierarchical level is larger for the 15-2 than for the 13-4. The evolution of
$r_{link}$ shows that when the coupling $\lambda$ is increased the number of
links synchronized in the $13-4$ network becomes larger than in the $15-2$
structure revealing that complete synchronization is then favored by the
presence of more external links connecting the first level communities.

\begin{figure}[!t]
\epsfig{file=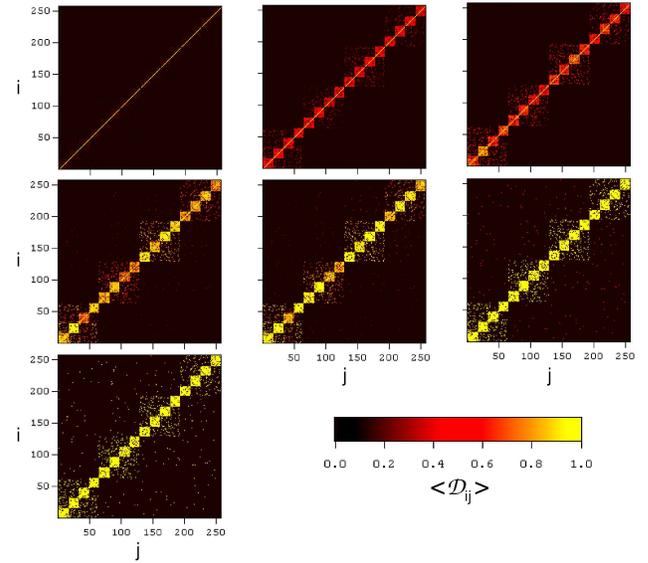,width=\columnwidth,angle=0,clip=1}
  \caption{(color online) We represent the degree of synchronization
  between pairs of connected nodes for several values of the coupling
  $\lambda$ in a $13-4$ modular network (with two organizational
  levels) of $N=256$ nodes. The color code denotes the value of the
  averaged (over different initial conditions) filtered matrix
  $\langle{\cal D}_{ij}\rangle\in [0,1]$.  The values of the coupling
  are (from left to right and top to bottom) $\lambda =0.011$,
  $0.026$, $0.032$, $0.035$, $0.038$, $0.046$, $0.210$ (corresponding
  to full synchronization). The pictures show that the order of
  synchronization is given by the organizational levels. The first
  community level is the first one to get synchronized, subsequently,
  second level nodes attain synchronization for a larger value of
  $\lambda$ and finally the full synchronized state is reached when
  outer links have $\langle D_{ij}\rangle=1$.}
\label{13-4}
\end{figure}

Fig.\ref{gc_comm} shows the size of the giant component of
synchronized clusters and the number of them as a function of
$\lambda$. An interesting effect of the community structure of the
networks and of the dynamics of the synchronization process is
revealed in the figure. Right at the value of $\lambda$ where the
onset of global coherence takes place, the size of $GC$ suddenly falls
to increase again at larger values of the coupling
strength. Additionally, note that this point coincides with that
corresponding to a change in the concavity of the $r_{link}(\lambda)$
curves. This change at the microscopic level is due to the
readjustment of links that connect synchronized nodes. In fact, as
Fig.\ \ref{nlinks} illustrates for both networks, in this region of
$\lambda$ values, the number of links connecting synchronized nodes of
the third level decreases while the number of those ascribed to the
second level raises. That is, the synchronization process takes place
in such a way that the first to synchronize are the nodes of the inner
community level, then the second and so on until the whole network
gets synchronized. The relevant fact is that in order for $r_{link}$
and $r$ to grow, the nodes and links of the second level adjust their
phases at the expense of those of the outer layer, the third
level. This is also reflected in the number of clusters of
synchronized links ($N_c$), i.e., the network appears like if the
nodes of the third level were ``temporarily'' disconnected. Moreover,
as the $13-4$ network has more links connecting the first and the
second hierarchical levels, $N_{links2}$ raises faster in this network
than in the $15-2$.

We have further inspected the synchronization path in modular
networks. This can be easily done and visualized by the representation
of the filtered matrix ${\cal D}$. It implies to reassign the values
of matrix ${\cal D}$ so that ${\cal D}_{ij}=1$ if ${\cal D}_{ij}>T$ ,
and ${\cal D}_{ij}=0$ otherwise. Plotting this filtered matrix for
different values of the coupling $\lambda$ one can easily determine
which links are the first to synchronize since the form of the
adjacency matrix (that includes all the physical links between nodes)
is also easy to interpret because of its nested
structure. Fig.\ref{13-4} shows how the community structure determines
the internal organization of the system in the route towards full
synchronization for the $13-4$ network. For this study we have
computed the value of the filtered matrix ${\cal D}$ for a number of
initial conditions and then took its average value so that $\langle
{\cal D}_{ij} \rangle\in [0,1]$ accounts for the synchronization
strength of the network link $(i,j)$. The results point out that link
synchronization depends on the organizational level they belong
to. Those connecting nodes belonging to the same first level community
are the fastest (in terms of the coupling strength $\lambda$) to reach
full synchronization. For larger values of $\lambda$ full
synchronization is attained progressively for the subsequent
organizational levels. Then, one can conclude that the inner the link
is the faster it gets synchronized in agreement with previous studies
reported above \cite{ArenasPRL06}.

\section{Conclusions}

In this paper we have explored several issues about synchronization in
complex networks of Kuramoto phase oscillators. Our main concern has
been the study of the synchronization patterns that emerge as the
coupling between non-identical oscillators increases. We have
described the degree of synchronization between each pair of connected
oscillators. The use of a new parameter, $r_{link}$, allows to
reconstruct the synchronization clusters from the dynamical data. We
have studied how the underlying topology (ranging from homogeneous to
heterogeneous structures) affects the evolution of synchronization
patterns. The results reveal that the route towards full
synchronization depends strongly on whether one deals with homogeneous
or heterogenous topologies. In particular, it has been shown that a
giant cluster of synchronization in heterogeneous networks comes out
from a unique core formed by highly connected nodes (hubs) whereas for
homogeneous networks several synchronization clusters of similar size
can coexist. In the latter case, a coalescence of these clusters is
observed in the synchronization path which is macroscopically
manifested by the sudden growth of global coherence. Another important
effect of the underlying topology is manifested in an anticipated
onset of global coherence for heterogeneous networks with respect to
more homogeneous topologies. However, the latter reaches the state of
full synchronization at lower values of the coupling strength,
therefore showing that statements about synchronizability of complex
networks are relative to the region of the phase diagram where they
operate. Additionally, we have shown that these systems are seen to
organize towards synchronization even when no macroscopic signs of
global coherence is observed.

Finally, the framework of structured networks has provided a useful
benchmark for testing the validity of the new parameter $r_{link}$ and
the information obtained from the computation of matrix ${\cal
D}$. The results obtained by means of these quantities allow to
conclude that for modular networks synchronization is first locally
attained at the most internal level of organization and, as the
coupling is increased, it progressively evolves toward outer shells of
the network. The latter process is however achieved at the expense of
partially readjusting some pairs of synchronized nodes between the
inner and outer community levels. Besides, we have obtained evidences
that a high cohesion at the first level communities produce a high
degree of local synchronization although it delays the appearance of
the global coherent state.

This study has extended the previous findings about the paths towards
synchronization in complex networks \cite{prljya}, and provides a
deeper understanding of phase synchronization phenomena on top of
complex topologies. In general, the work supports the idea that in the
absence of analytical tools to confront the resolution of non-linear
dynamical models in complex networks, the introduction of new
parameters to describe the statistical properties of the emergence of
local patterns is needed as they give novel and useful information
that might guide our comprehension of these phenomena. On more general
grounds, this work adds to other recent findings \cite{chaos,games}
about the topology emerging from dynamical processes. The evidences
that are being accumulated point to a dynamical organization, both at
the local and global scales, that is driven by the underlying
topology. Whether or not this intriguing regularity has something to
do with the ubiquity of complex heterogeneous networks in Nature is
not clear yet. More works in this direction are needed, but we think
that they may ultimately lead to uncover important universal relations
between the structure and function of complex natural systems that
form networks. Another issue to explore in future works concerns the
behavior of non-linear dynamical systems on top of directed networks
\cite{timme}, which will allow deeper insights into the behavior of
natural systems.

\begin{acknowledgments}
  We thank J.A. Acebr\'on, S. Boccaletti, A. D\'{\i}az-Guilera,
  C.J. P\'{e}rez-Vicente and V. Latora for helpful
  comments. J.G.G. and Y.M. are supported by MEC through a FPU grant
  and the Ram\'{o}n y Cajal Program, respectively. This work has been
  partially supported by the Spanish DGICYT Projects
  FIS2006-13321-C02-02, FIS2006-12781-C02-01 and FIS2005-00337 and by
  the European NEST Pathfinder project GABA under contract 043309.
\end{acknowledgments}


\begin{thebibliography}{27}
\bibitem{note1} Note that this is only possible in networks with power-law
degree distributions, but with a very small probability as $P(k)\sim
k^{-\gamma}$ with $\gamma>0$. In these cases, mean-field solutions
independent on $N$ are recovered, with slight differences in the onset
of synchronization of all-to-all and scale-free networks
\cite{RestrepoPRE05}.
\bibitem{note2} A direct comparison with the all-to-all
  Kuramoto model is difficult to stablish, since one system is
  extensive (SF networks) and the other depends on $N$ (KM) so that
  the results does not corresponds to the same effective
  coupling.
\bibitem{note3} Note that the main difference between both measures is
that one refers to the degree of synchronization of nodes ($r$) with
respect to the average phase and the other ($r_{link}$) to the degree of
synchronization between every pair of connected nodes.
\bibitem{note4} Note that the size of the network is relatively small
($N=100$) and thus the critical point is shifted to lower values
($\lambda_c^{ER}\simeq 0.07$ in this case) than that found using a FSS
analysis.
\end{thebibliography}
\end{document}